\documentclass[a4paper,12pt]{article}
\pdfoutput=1 

\usepackage{jheppub} 
\usepackage[T1]{fontenc} 

\usepackage[latin1]{inputenc}
\usepackage[english]{babel}

\usepackage{amssymb,amsthm,cancel,hyperref,graphicx,xcolor}
\usepackage{picinpar,graphicx,xy}
\usepackage[subnum]{cases}
\usepackage{booktabs}
\usepackage{mathrsfs}
\usepackage{amsfonts}
\usepackage{latexsym}
\usepackage{booktabs,graphicx,hyperref,epsfig,multirow}
\usepackage{bm}
\usepackage{comment}
\allowdisplaybreaks
\def\bea{\begin{align}}
\def\eea{\end{align}}
\def\beq{\begin{equation}}
\def\eeq{\end{equation}}
\def\ba{\begin{eqnarray}}
\def\ea{\end{eqnarray}}
\def\be{\begin{equation}}
\def\ee{\end{equation}}
\definecolor{darkgreen}{HTML}{008000}
\newcommand{\sss}{\scriptscriptstyle\rm}
\newcommand{\muf}{\mu_{\rm\sss F}}

\newcommand{\mH}{m_{\sss H}}

\newcommand{\abs}[1]{\left|\,#1\,\right|}

\newcommand{\Ord}{\mathcal{O}}

\newcommand{\as}{\alpha_s}
\newcommand{\barkt}{\bar{k}_{\sss T}}

\def\({\left(}
\def\){\right)}
\def\[{\left[}
\def\]{\right]}
\def    \hepph  #1 {{\tt hep-ph/#1}}
\def    \hepex  #1 {{\tt hep-ex/#1}}
\long\def\symbolfootnote[#1]#2{\begingroup%
\def\thefootnote{\fnsymbol{footnote}}\footnote[#1]{#2}\endgroup}

\numberwithin{equation}{section}
\def\lapprox{\lower .7ex\hbox{$\;\stackrel{\textstyle <}{\sim}\;$}}
\def\gapprox{\lower .7ex\hbox{$\;\stackrel{\textstyle >}{\sim}\;$}}

\renewcommand{\(}{\left(}
\renewcommand{\)}{\right)}

\newcommand{\MSbar}{$\overline{\rm MS}$}

\newcommand{\plusq}[1]{\left[#1\right]_+}
\newcommand{\Ca}{C_{\rm\sss A}}
\newcommand{\Cf}{C_{\rm\sss F}}

\newcommand{\kt}{k_{\rm\sss T}}
\newcommand{\pt}{p_{\rm\sss T}}

\definecolor{darkblue}{rgb}{0,0,0.7}

\begin{document}
\begin{flushleft}
\begin{figure}[h]
\includegraphics[width=.2\textwidth]{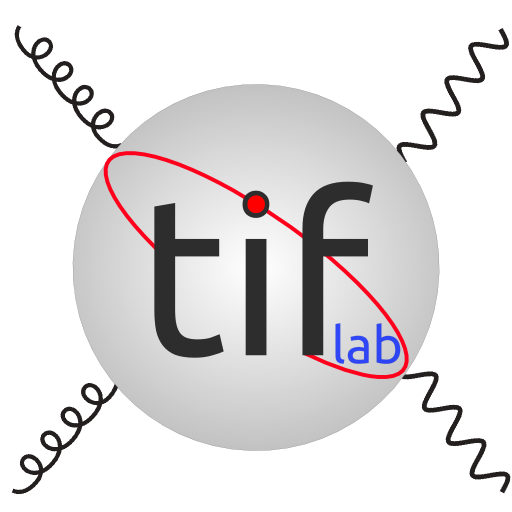}
\end{figure}
\end{flushleft}
\vspace{-5.0cm}
\begin{flushright}
TIF-UNIMI-2017-10
\end{flushright}

\vspace{2.0cm}

\begin{center}
{\Large \bf 
Double Differential High Energy Resummation}
\end{center}


\begin{center}
Claudio Muselli$^1$. \\
\vspace{.3cm}
{\it
{}$^1$Tif Lab, Dipartimento di Fisica, Universit\`a di Milano and\\
INFN, Sezione di Milano, Via Celoria 16, I-20133 Milano, Italy
}
\end{center}

\begin{center}
{\bf \large Abstract}
\end{center}
In this paper we study a general framework to perform leading log high energy resummation with complete dependence from the rapidity and the transverse momentum of the studied system. As an application, high energy resummed expression for the Higgs boson production in the effective field theory framework will be derived. Explicit result for this process will be cross-checked against fixed order evaluations up to NNLO. Consistency with known high energy resummation for single differential rapidity or transverse momentum distribution, as well as for inclusive cross section, will also be shown.

\clearpage
\tableofcontents
\clearpage

\section{Introduction}
In recent years, great improvements have been made in the context of high energy resummation. Even if the general accuracy is still leading log (LL$x$ for short), original theory derived for inclusive cross sections in the first pioneering works~\cite{Catani:1990xk,Catani:1990eg,Catani:1994sq} and successfully applied  to various relevant processes~\cite{Ball:2001pq,Hautmann:2002tu,Marzani:2008az,Marzani:2008uh,Diana:2009xv,Zoia:2017vfn}, was first extended to single rapidity distributions in Ref.~\cite{Caola:2010kv} and then to single transverse momentum distributions in Ref.~\cite{Forte:2015gve}. Moreover in Ref.~\cite{Zoia:2017vfn} a first application to coloured final state was presented opening the possibility to study with this technique also jet observables. 

Further improvements appear in recent years also in the context of the numerical evaluation of the  resummed results. Indeed, it is well known~\cite{Catani:1990eg,Catani:1994sq,Caola:2010kv} that high energy resummation of collider observables is based on the LL$x$ resummation of DGLAP anomalous dimensions and splitting functions. Such a resummation is far to be trivial since several formally NLL$x$ effects must be inserted as well to reach a final stable LL$x$ resummed result~\cite{AltarelliBallForte,Altarelli:1999vw,Altarelli:2001ji}. Even if theoretical procedure was known since a long time~\cite{Altarelli:2001ji}, a practical implementation of a matched NNLO+NLL$x$ resummed DGLAP anomalous dimension has been reached only recently~\cite{Bonvini:2016wki,Bonvini:2017ogt}, thanks to some clever modifications of the original technique. 

Even from the phenomenological point of view, high energy resummation reveals its importance at LHC nowadays. First in Ref.~\cite{Caola:2010kv}, high energy resummation was used to study the impact of mass quark effects on the Higgs $\pt$ spectrum revealing important information about its all order structure and presenting a possible approximation for the unknown NLO coefficient. Moreover, very recently~\cite{Ball:2017otu} a full LL$x$ resummed PDFs set was derived by the NNPDF collaboration by supplying fixed order calculations and DGLAP evolution with resummation effects. This has significantly increased the theoretical accuracy of the PDF fits in the relevant kinematical region~\cite{Rottoli:2017ifw, Ball:2017otu}.

This paper stands as the seamless continuation of this process. We will unify the high energy resummation for single differential rapidity distributions of Ref.~\cite{Caola:2010kv} and the high energy resummation for single differential transverse momentum distributions of Ref.~\cite{Forte:2015gve} in a unique framework, performing a double differential high energy resummation for our studied system. The general theoretical discussion we are going to present will be valid for any desired final state, both colourless than coloured.

This is not a mere theoretical exercise since contemporary knowledge of rapidity and transverse momentum for a studied final system is fundamental in many phenomenological applications. In the context of singlet boson production a double differential prediction is needed in order to implement even in the resummed result the experimental cuts which are present in all the LHC analysis. Moreover, the resummation of any jet observable relies on the application of a particular jet cluster algorithms. In almost all of them, an exclusive knowledge of rapidity, transverse momentum and azimuthal angle is needed to correctly cluster final partons into jets. 

As an application, in this paper we will concentrate on the Higgs boson production in the effective field theory context. Its simplicity permits us to perform analytically some consistency checks on our final result. Its expansion will be cross-checked against fixed order evaluations up to NNLO using the technique of Ref.~\cite{Caola:2010kv, Muselli:2017ikh}; moreover, single differential and inclusive high energy resummations will be derived as a by-product performing suitable rapidity or/and transverse momentum integrations.

The paper is organized as follows. In Sec~\ref{sec:general}, general treatment about high energy resummation using generalized ladder expansion is briefly reviewed for a generic final state observable. Then in Sec.~\ref{sec:doublediff}, the particular case of double differential cross section will be treated in details, presenting our final resummed expression. Finally in Sec.~\ref{sec:Higgshigh} the case of Higgs boson production in the effective field theory is studied and high energy resummation is performed explicitly at double differential level. In the same section, consistency of this result is shown with high energy predictions of Refs.~\cite{Caola:2010kv,Forte:2015gve,Hautmann:2002tu} for less exclusive observables. Moreover, in Sec.~\ref{subsec:check} we will perform a further cross-check against fixed order evaluations up to NNLO. At the end, conclusions and outlooks are drawn in Sec.~\ref{sec:concl}. 

\section{High Energy Resummation for a generic observable}
\label{sec:general}

In this section we are going to review the basis of high energy resummation for a generic observable at LL$x$. At the end of the section we will specify better the type of observable we are considering but for now all the discussion is completely general.

Let us consider in order to fix the notation a general gluon fusion process, such as Higgs boson production:
\beq
g\(p\) + g\(n\) \to \mathcal{S}\(p_{\mathcal{S}}\) + X
\eeq
where $\mathcal{S}$ is the studied system and it could be colourless or coloured while $X$ is any extra radiation. We are going to call $Q$ the hard scale associated to the system $\mathcal{S}$ and $\sqrt{s}$ the partonic centre-of-mass energy.

\begin{figure}[htb!]
\centering
\includegraphics[width=0.6\linewidth]{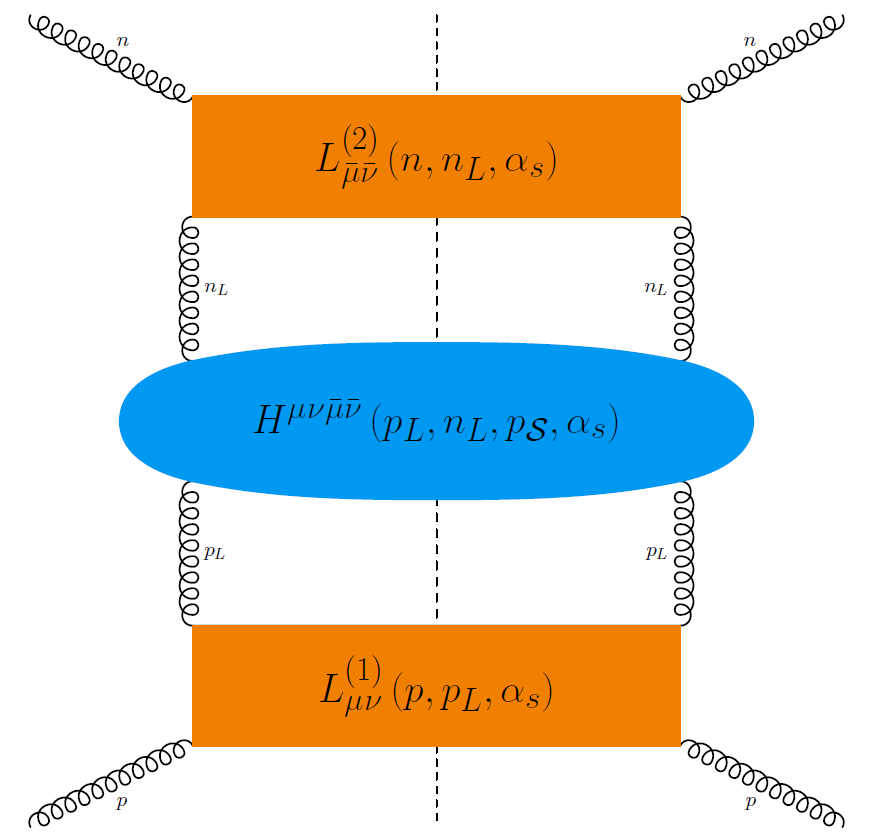}
\caption{Hard-ladders decomposition of a general partonic observable in $\kt$-factorization}
\label{fig:hardladder}
\end{figure}

The basis of any high energy resummation approach is $\kt$-factorization. \\
$\kt$-factorization states that at high energy leading contributions to any partonic observable come from cut diagrams which are at least two-gluon-irreducible (2GI) in the $t$-channel. Moreover, using power counting arguments it can be proved~\cite{Ellis:1978ty} that any radiation connecting the two initial legs is suppressed by powers of the centre-of-mass energy $s$. The result is that any (dimensionless) infra-red and collinear safe partonic observable $\mathcal{\hat{O}}$ can be written in terms of a process dependent \emph{hard part} $H^{\mu\nu\bar{\mu}\bar{\nu}}$ and two universal \emph{ladders} of emission $L_{\mu\nu}$, $L_{\bar{\mu}\bar{\nu}}$. Situation is depicted in Fig.~\ref{fig:hardladder}. At LL$x$ only the longitudinal tensor components of $H$ and $L$ survives and we can thus write in formula:
\begin{align}
\label{eq:ktfact2}
\mathcal{\hat{O}}\(\hat{\tau},\{v\},\frac{\muf^2}{Q^2},\as\)&=\int\left[ \frac{\hat{\tau}}{2 z \bar{z}} H_{\parallel, \parallel}\(\frac{\hat{\tau}}{z\bar{z}},\frac{\kt^2}{Q^2},\frac{\barkt^2}{Q^2},\{v\},\as\)\right]\notag\\
&\left[2\pi L_{\parallel}^{(1)}\(\frac{\muf^2}{\kt^2},\as\)\right]\left[2\pi L_{\parallel}^{(2)}\(\frac{\muf^2}{\kt^2},\as\)\right]\notag\\
&\frac{dz}{z}\frac{d\bar{z}}{\bar{z}}\frac{d\kt^2}{\kt^2}\frac{d\barkt^2}{\barkt^2} \frac{d\theta}{2\pi}\frac{d\bar{\theta}}{2\pi}+ \Ord\(z,\bar{z}\)
\end{align}
where we have defined 
\beq
\label{eq:taudef}
\hat{\tau}=\frac{Q^2}{s},
\eeq
the following kinematics for the incoming off-shell gluons (see Fig.~\ref{fig:hardladder})
\begin{subequations}
\label{eq:pLnL}
\begin{align}
p_L &= z p + \mathbf{k} &  \abs{\mathbf{k}}^2&=-\kt^2,\\
n_L &= \bar{z} n + \mathbf{\bar{k}} & \abs{\mathbf{\bar{k}}}^2 &= -\barkt^2,
\end{align}
\end{subequations}
$\theta$ and $\bar{\theta}$ as the azimuthal angles of the transverse momenta $\mathbf{k}$ and $\mathbf{\bar{k}}$ and $\{v\}$ as the set of variables which characterizes the desired observable. Clearly the limit we are interested in is $\hat{\tau} \to 0$ and we are supposing to resum leading logarithmic contributions of type $\as^k \ln^k\hat{\tau}$ inside our partonic observable. Moreover delta constraint such as
\beq
\delta\(\{v\}-\{v\(p_{\mathcal{S}}\)\}\),
\eeq
which denotes the particular partonic observable we choose in Eq.~\eqref{eq:ktfact2}, is supposed to be included in the hard part. 

We can define a process-dependent hard coefficient function $C_{\mathcal{\hat{O}}}$ as follows
\begin{align}
\label{eq:coeffunction}
C_{\mathcal{\hat{O}}}\(\frac{\hat{\tau}}{z \bar{z}},\frac{\kt^2}{Q^2},\frac{\barkt^2}{Q^2}, \{v\},\as\) &\equiv \int \frac{d\theta}{2\pi} \frac{d\bar{\theta}}{2\pi} \frac{\hat{\tau}}{2 z \bar{z}} H_{\parallel, \parallel}\(\frac{\hat{\tau}}{z\bar{z}},\frac{\kt^2}{Q^2},\frac{\barkt^2}{Q^2},\as\)\notag\\
&\qquad\qquad\qquad\qquad\delta\(v_i -v_i\(p_L,n_L,q_1,\dots,q_n\)\)\notag\\
&\equiv \int \frac{d\theta}{2\pi} \frac{d\bar{\theta}}{2\pi} \frac{\hat{\tau}}{2 z \bar{z}}\left[\mathcal{P}^{\mu\nu} \mathcal{P}^{\bar{\mu}\bar{\nu}} H_{\mu\nu\bar{\mu}\bar{\nu}}\right]\notag\\
&\qquad\qquad\qquad\qquad\delta\(v_i -v_i\(p_L,n_L,q_1,\dots,q_n\)\),
\end{align}
where projectors $\mathcal{P}$ are defined as
\begin{align}
\label{eq:proj}
\mathcal{P}^{\mu\nu}&=\frac{\kt^{\mu} \kt^{\nu}}{\kt^2} & \mathcal{P}^{\bar{\mu}\bar{\nu}}&=\frac{\barkt^{\bar{\mu}} \barkt^{\bar{\nu}}}{\barkt^2}.
\end{align}
and we decide to explicit the delta constraint contained in the hard part.

The coefficient function Eq.~\eqref{eq:coeffunction} owns a simple physical interpretation. It represents the observable $\mathcal{\hat{O}}$ evaluated for the LO off-shell process
\beq
g^* + g^* \to \mathcal{S}
\eeq
where projectors Eq.~\eqref{eq:proj} can be viewed as a polarization sum prescription for the two incoming off-shell gluons. In Ref.~\cite{Zoia:2017vfn} the authors showed that coefficient function Eq.~\eqref{eq:coeffunction} can be computed using the LO off-shell process even if its on-shell counterpart is zero or trivial; resummation prescription still works.

Using definition of the hard coefficient function Eq.~\eqref{eq:coeffunction} into Eq.~\eqref{eq:ktfact2} we rewrite it as
\begin{align}
\label{eq:sigma1}
\mathcal{\hat{O}}\(\hat{\tau},\{v\},\frac{\muf^2}{Q^2},\as;\epsilon\)&=\(\muf\)^{2\epsilon}\int
C_{\mathcal{\hat{O}}}\(\frac{\hat{\tau}}{z
  \bar{z}},\frac{\kt^2}{Q^2},\frac{\barkt^2}{Q^2},\{v\},\as;
\epsilon\)\notag\\
&\left[2\pi L^{(1)}_{\parallel}\(z,
  \(\frac{\muf^2}{\kt^2}\)^\epsilon,\as; \epsilon\)\right]\left[2\pi
  L^{(2)}_{\parallel}\(\bar{z},\(\frac{\muf^2}{\barkt^2}\)^\epsilon,\as;
  \epsilon\)\right]\notag\\
  &\frac{dz}{z}\frac{d\bar{z}}{\bar{z}}\frac{d\kt^2}{\(\kt^2\)^{1+\epsilon}}\frac{d\barkt^2}{\(\barkt^2\)^{1+\epsilon}}.
\end{align}
Now we are ready to perform the LL$x$ resummation by computing the ladders using generalized ladder expansion approach of Refs.~\cite{Curci:1980uw,Caola:2010kv,Forte:2015gve}. Before starting however, a remark about renormalization and factorization scale dependence of Eq.~\eqref{eq:sigma1} is due. 

First of all, running coupling effects are in the context of high energy resummation beyond our working accuracy. For this reason we are working in all the expression of this and following sections  at fixed coupling $\as$. Different is the situation of factorization scale. 

In principle our factorized expression Eq.~\eqref{eq:sigma1} shows IR divergences both in the hard part and in the ladder part. Therefore we write previous equation in $d=4-2\epsilon$ dimensions, using dimensional regularization to identify all the singularities. Now, we can limit our discussion to cases where the hard part is 2PI rather than 2GI and thus free of collinear singularities. The extension to the case where the hard part is not finite (as in the DY case) is straightforward even if not totally trivial and we refer to Ref.~\cite{Marzani:2008uh} for further details. Instead, IR singularities into the ladders are going to be iterative subtracted using the generalized ladder expansion of Refs.~\cite{Curci:1980uw,Caola:2010kv,Forte:2015gve}; from this subtraction we are going to obtain our final LL$x$ resummed expression. 

In the generalized ladder approach, ladders are viewed as multiple insertions of a proper LL$x$ emission kernel $K$, expressed in $d$ dimensions. We require our result to be finite after each insertion of $K$, thus iterative subtracting divergences into renormalized PDFs.

However, this procedure has to be performed differently according to the type of partonic observable we are studying. In particular, we have to distinguish between two cases:
\begin{itemize}
\item observables whose value is independent from the number of emissions in the ladders. Ladders emissions can thus be treated as independent one from each other.
\item observable whose value is dependent from the number of emissions in the ladders. Ladders are then correlated with the hard part and they can not be resummed directly.
\end{itemize}
In the first case iterative subtraction works as in the inclusive cross section case; the only modification is the different definition for the hard coefficient function Eq.~\eqref{eq:coeffunction} (see Ref.~\cite{Forte:2015gve} for details). Examples of this type are transverse momentum distributions, invariant mass distributions or some jet observables such as one-jet inclusive cross sections or leading-$\pt$ jet distributions. On the contrary, in the second case, particular extra integral transforms are necessary to factorize the delta constraint contained in the definition of the hard part to disentangle it from the number of emissions. Rapidity distributions or double differential distributions are included in this case. For this reason, in the next section, we are going to focus on the double differential rapidity and transverse momentum distribution and we present a suitable integral transform which works in this case. In this way we are able to reach our desired LL$x$ resummed expression.

\subsection{Double Differential High Energy Resummation}
\label{sec:doublediff}

Our aim is to resum the LL$x$ contributions for the double differential distribution with respect to rapidity and transverse momentum. In this case, the generic hard coefficient function Eq.~\eqref{eq:coeffunction} has to be defined as
\begin{multline}
C_{\pt, y}\(\frac{\hat{\tau}}{z
  \bar{z}},\xi,\bar{\xi},\xi_p, y ,\as;
\epsilon\)\equiv \int \frac{d\theta}{2\pi} \frac{d\bar{\theta}}{2\pi} \frac{\hat{\tau}}{2 z \bar{z}}\left[\mathcal{P}^{\mu\nu} \mathcal{P}^{\bar{\mu}\bar{\nu}} H_{\mu\nu\bar{\mu}\bar{\nu}}\right]\\
\delta\(\xi_p-\xi-\bar{\xi}-2\sqrt{\xi \bar{\xi}}\cos\theta\)\delta\(y-\frac{1}{2}\ln\frac{z}{\bar{z}}\),
\label{eq:Cpty}
\end{multline}
with the following dimensionless ratios
\begin{align}
\label{eq:xipdef}
\xi_p&=\frac{\pt^2}{Q^2} & \xi&=\frac{\kt^2}{Q^2} & \bar{\xi}&=\frac{\barkt^2}{Q^2},
\end{align}
$\pt$ the transverse momentum of the final studied system and  $z$, $\bar{z}$, $\kt^2$ and $\barkt^2$ given in Eq.~\eqref{eq:pLnL}.

\begin{figure}[htb]
\centering
\includegraphics[width=0.8\textwidth]{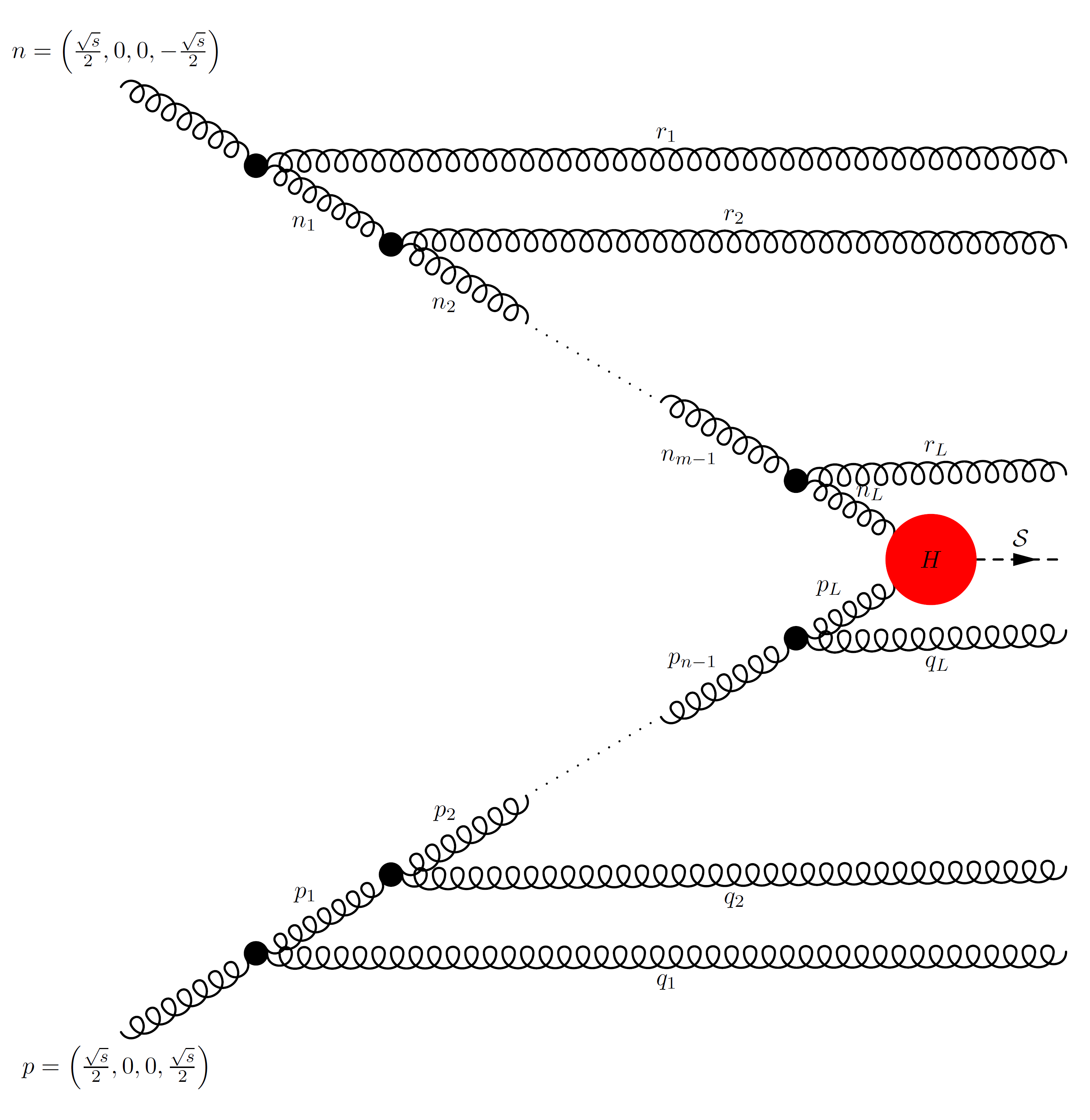}
\caption{Kinematics of the ladder}
\label{fig:kinematics}
\end{figure}

By inserting Eq.~\eqref{eq:Cpty} into general $\kt$ factorized expression Eq.~\eqref{eq:sigma1} we obtain that high energy resummation for $\frac{d\sigma}{dy d\xi_p}$ is given by:
\begin{align}
\label{eq:doublediffsigma1}
\frac{d\sigma}{d\xi_p dy}\(\hat{\tau},\xi_p, y ,\frac{\muf^2}{Q^2},\as;\epsilon\)&=\(\muf\)^{2\epsilon}\int
C_{\pt, y}\(\frac{\hat{\tau}}{z
  \bar{z}},\frac{\kt^2}{Q^2},\frac{\barkt^2}{Q^2},\xi_p, y ,\as;
\epsilon\)\notag\\
&\left[2\pi L^{(1)}_{\parallel}\(z,
  \(\frac{\muf^2}{\kt^2}\)^\epsilon,\as; \epsilon\)\right]\left[2\pi
  L^{(2)}_{\parallel}\(\bar{z},\(\frac{\muf^2}{\barkt^2}\)^\epsilon,\as;
  \epsilon\)\right]\notag\\
  &\frac{dz}{z}\frac{d\bar{z}}{\bar{z}}\frac{d\kt^2}{\(\kt^2\)^{1+\epsilon}}\frac{d\barkt^2}{\(\barkt^2\)^{1+\epsilon}}. 
  \end{align}
  
By inspecting general ladder kinematics, which is given by (see also Fig.~\ref{fig:kinematics})
\begin{subequations}
\label{eq:kinematics}
\begin{align}
p_1&=z_1 p - \mathbf{k}_1\\
q_1&=\(1-z_1\) p + \mathbf{k}_1\\
p_2&=z_2 z_1 p - \mathbf{k}_2 \\
q_2&=\(1-z_2\)z_1 p + \mathbf{k}_2 - \mathbf{k}_1\\
&\dots\dots\dots\\
p_L&=z_1 \dots z_n p - \mathbf{k} \\
q_L&=\(1-z_n\)z_1 \dots z_{n-1} p +\mathbf{k} - \mathbf{k}_{n-1}\\
\quad\notag\\
n_1&=\bar{z}_1 p - \mathbf{\bar{k}}_1\\
r_1&=\(1-\bar{z}_1\) p + \mathbf{\bar{k}}_1\\
n_2&=\bar{z}_2 \bar{z}_1 p - \mathbf{\bar{k}}_2 \\
r_2&=\(1-\bar{z}_2\)\bar{z}_1 p +\mathbf{\bar{k}}_2 - \mathbf{\bar{k}}_1\\
&\dots\dots\dots\\
n_L&=\bar{z}_1\dots \bar{z}_m n - \mathbf{\bar{k}} \\
r_L&=\(1-\bar{z}_m\)\bar{z}_1 \dots \bar{z}_{m-1} n +\mathbf{\bar{k}}- \mathbf{\bar{k}}_{m-1},
\end{align}
\end{subequations}
you can easily convince yourself that, after multiple kernel insertions, the particular value of $z$, $\bar{z}$ contained in the rapidity delta constraint in Eq.~\eqref{eq:doublediffsigma1} does depend on the number $n$ and $m$ of insertions in each leg. Indeed, we have following relations
\begin{align}
z&=z_1 \dots z_n \\
\bar{z}&=\bar{z}_1 \dots \bar{z}_m
\end{align}
and then ladder resummation can not be performed in exactly the same way as in total cross section case. A similar situation was already solved in the case of single rapidity distribution in Ref.~\cite{Caola:2010kv} and here we are going to present a analogue derivation.

\begin{figure}[hb!]
\centering
\includegraphics[width=0.6\linewidth]{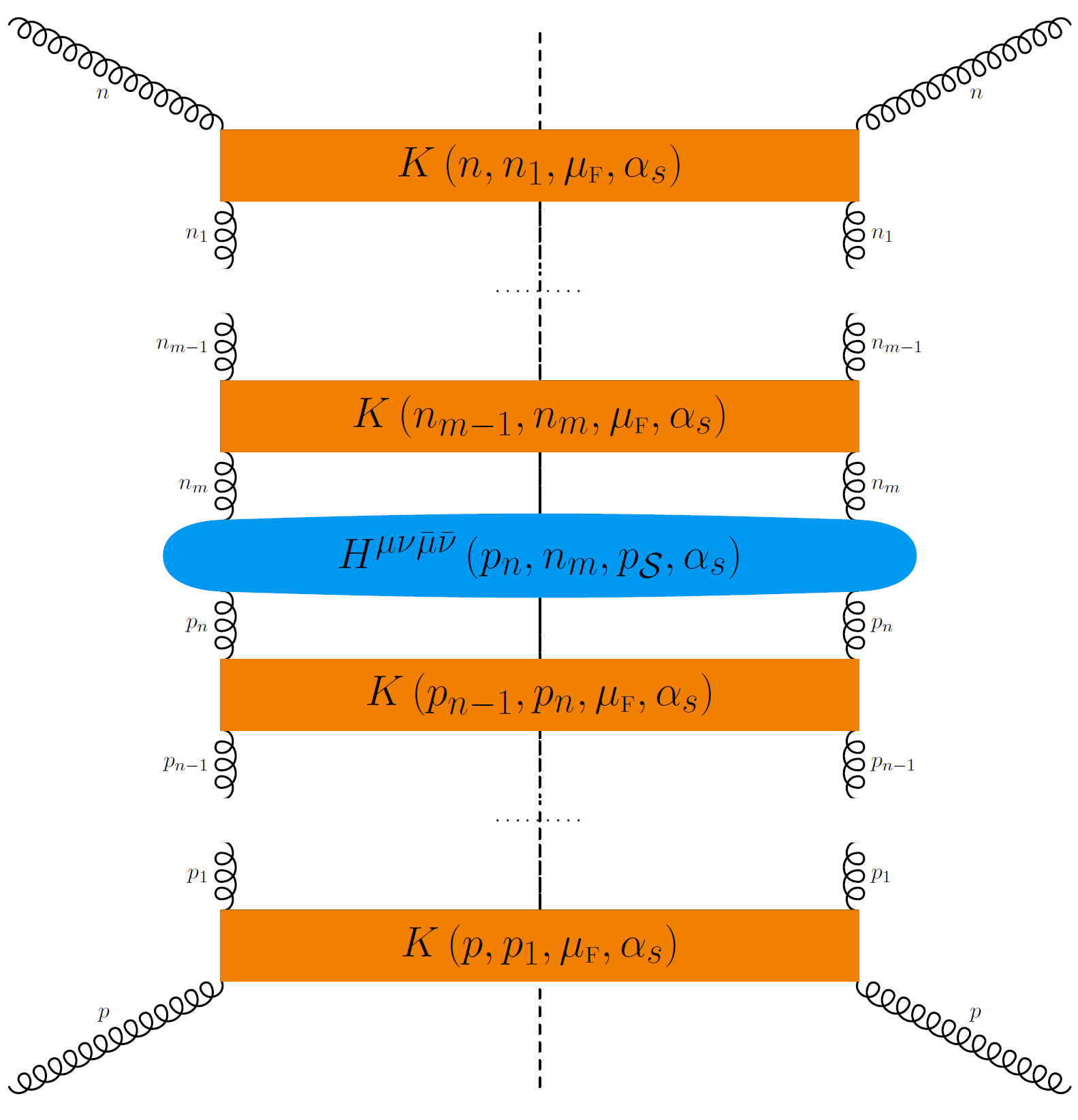}
\caption{Schematically computation of ladders as multiple insertion of the Kernel $K$}
\label{fig:kernels}
\end{figure}

First, we expand at LL$x$ ladders $L^{(1,2)}$ of Eq.~\eqref{eq:doublediffsigma1} through multiple insertions of a proper kernel $K$. Procedure is depicted in Fig.~\ref{fig:kernels}. We thus write:
\begin{align}
\label{eq:doublediffsigma2}
&\frac{d\hat{\sigma}^{n,m}}{dy d\xi_p}\(\hat{\tau},\xi_p,y,\frac{\muf^2}{Q^2},\as;\epsilon\)=\int_0^1 dz_n\,\int_0^\infty \left[K\(z_n,\(\frac{\muf^2}{Q^2\xi_n}\)^\epsilon,\as;\epsilon\)\right]\frac{d\xi_n}{\xi_n^{1+\epsilon}}\times\notag\\
&\times\int_0^1 d\bar{z}_m\,\int_0^{\infty}\left[K\(\bar{z}_m,\(\frac{\muf^2}{Q^2\xi_{m}}\)^\epsilon,\as;\epsilon\)\right]\frac{d\bar{\xi}_m}{\bar{\xi}_{m}^{1+\epsilon}} \notag\\
&C_{\pt,y}\(\frac{\hat{\tau}}{z
  \bar{z}},\xi,\bar{\xi},\xi_p, y ,\as;
\epsilon\)\delta\(\frac{1}{2}\ln\frac{z}{\bar{z}}-\frac{1}{2}\ln\frac{z_1 \dots z_n}{\bar{z}_1 \dots \bar{z}_m}\)\times\notag\\
&\times \int_0^1 dz_{n-1}\,\int_0^{\xi_n} \left[K\(z_{n-1},\(\frac{\muf^2}{Q^2 \xi_{n-1}}\)^{\epsilon},\as;\epsilon\)\right]\frac{d\xi_{n-1}}{\xi_{n-1}^{1+\epsilon}}\notag\\
&\times \dots \times \int_0^1 dz_1\,\int_0^{\xi_2}\left[K\(z_1,\(\frac{\muf^2}{Q^2 \xi_1}\)^\epsilon,\as;\epsilon\)\right]\frac{d\xi_1}{\xi_1^{1+\epsilon}}\times\notag\\
&\times\int_0^1 d\bar{z}_{m-1}\,\int_0^{\bar{\xi}_m}\left[K\(\bar{z}_{m-1},\(\frac{\muf^2}{Q^2\xi_{m-1}}\)^\epsilon,\as;\epsilon\)\right]\frac{d\bar{\xi}_{m-1}}{\bar{\xi}_{m-1}^{1+\epsilon}}\notag\\
&\times \dots \times \int_0^1 d\bar{z}_1\, \int_0^{\bar{\xi}_2}\left[K\(\bar{z}_{1},\(\frac{\muf^2}{Q^2 \xi_1}\)^\epsilon,\as;\epsilon\)\right]\frac{\bar{\xi}_1}{\bar{\xi}_1^{1+\epsilon}}
\end{align}
where we divide rapidity delta constraint
\beq
\delta\(y-\frac{1}{2}\ln\frac{z_1 \dots z_n}{\bar{z}_1 \dots \bar{z}_m}\)=\delta\(y-\frac{1}{2}\ln\frac{z}{\bar{z}}\)\delta\(\frac{1}{2}\ln\frac{z}{\bar{z}}-\frac{1}{2}\ln\frac{z_1 \dots z_n}{\bar{z}_1 \dots \bar{z}_m}\)
\eeq
into its hard and ladder parts. We include the hard part delta constraint into $C_{\pt ,y}$ definition in Eq.~\eqref{eq:doublediffsigma2}, according to general $C_{\mathcal{O}}$ definition, Eq.~\eqref{eq:coeffunction}.  

Now, the goal is reached by performing the right transform which factorizes ladder rapidity delta constraint. Indeed this is the only part which mixes ladders emissions throughout various $z_i$ and $\bar{z}_i$. By defining the following Mellin-Fourier Transform,
\begin{align}
\label{eq:Fourierdef}
\frac{d\hat{\sigma}^{n,m}}{dy d\xi_p}\(N,\xi_p,b,\frac{\muf^2}{Q^2},\as;\epsilon\)&=\int_0^1 d\hat{\tau}\, \hat{\tau}^{N-1}\int_{-\infty}^{\infty} dy\, e^{-i b y} \frac{d\hat{\sigma}^{n,m}}{dy d\xi_p}\(\hat{\tau},\xi_p,y,\frac{\muf^2}{Q^2},\as;\epsilon\)
\end{align}
we rewrite Eq.~\eqref{eq:doublediffsigma2} in the following factorized form
\begin{align}
\label{eq:doublediffsigma3}
&\frac{d\hat{\sigma}^{n,m}}{dy d\xi_p}\(N,\xi_p,b,\frac{\muf^2}{Q^2},\as;\epsilon\)=\int_0^\infty \left[\gamma\(N-\frac{i b}{2},\(\frac{\muf^2}{Q^2\xi_n}\)^\epsilon,\as;\epsilon\)\right]\frac{d\xi_n}{\xi_n^{1+\epsilon}}\times\notag\\
&\times\int_0^{\infty}\left[\gamma\(N+\frac{i b}{2},\(\frac{\muf^2}{Q^2\xi_{m}}\)^\epsilon,\as;\epsilon\)\right]\frac{d\bar{\xi}_m}{\bar{\xi}_{m}^{1+\epsilon}} C_{\pt,y}\(N,\xi_n,\bar{\xi}_m,\xi_p, b ,\as;
\epsilon\)\times\notag\\
&\times \int_0^{\xi_n} \left[\gamma\(N-\frac{i b}{2},\(\frac{\muf^2}{Q^2 \xi_{n-1}}\)^{\epsilon},\as;\epsilon\)\right]\frac{d\xi_{n-1}}{\xi_{n-1}^{1+\epsilon}}\notag\\
&\times \dots \times \int_0^{\xi_2}\left[\gamma\(N-\frac{i b}{2},\(\frac{\muf^2}{Q^2 \xi_1}\)^\epsilon,\as;\epsilon\)\right]\frac{d\xi_1}{\xi_1^{1+\epsilon}}\times\notag\\
&\times\int_0^{\bar{\xi}_m}\left[\gamma\(N+\frac{i b}{2},\(\frac{\muf^2}{Q^2\xi_{m-1}}\)^\epsilon,\as;\epsilon\)\right]\frac{d\bar{\xi}_{m-1}}{\bar{\xi}_{m-1}^{1+\epsilon}}\notag\\
&\times \dots \times\int_0^{\bar{\xi}_2}\left[\gamma\(N+\frac{i b}{2},\(\frac{\muf^2}{Q^2 \xi_1}\)^\epsilon,\as;\epsilon\)\right]\frac{\bar{\xi}_1}{\bar{\xi}_1^{1+\epsilon}}
\end{align}
where we define $C_{\pt,y}$ in Mellin-Fourier space as
\begin{align}
\label{eq:Cptyb}
C_{\pt, y}\(N,\xi,\bar{\xi},\xi_p,b,\as\)&=
&\int_0^1 dx\hat{\tau}\, \hat{\tau}^{N-1} \int_{-\infty}^{+\infty} dy\,e^{-i b y} C_{\pt,y}\(\hat{\tau},\xi,\bar{\xi},\xi_p,y,\as\).
\end{align}
In Eq.~\eqref{eq:doublediffsigma3} $\gamma$ stands for the LL$x$ DGLAP anomalous dimension in $d=4-2\epsilon$ dimensions; it is the Mellin transformed version of the generalized ladder kernel $K$.

It is interesting to observe that Fourier integration does not exit from the physical region. This is due to the fact that rapidity boundary
\beq
-y_{\rm max} < y < y_{\rm max}
\eeq
with
\beq
y_{\rm max}=\frac{1}{2}\ln\frac{1+\sqrt{1-\frac{4\hat{\tau}\(1+\xi_p\)}{\(1+\hat{\tau}\)^2}}}{1-\sqrt{1-\frac{4\hat{\tau}\(1+\xi_p\)}{\(1+\hat{\tau}\)^2}}}.
\eeq
tends to infinity when $\hat{\tau}$ approaches zero, thus in the high energy regime. 

Since now Eq.~\eqref{eq:doublediffsigma3} is factorized, following steps to reach high energy resummation can be performed exactly as in Refs.~\cite{Caola:2010kv,Forte:2015gve}. Therefore, we require Eq.~\eqref{eq:doublediffsigma3} to be finite after each $\xi_i$ or $\bar{\xi}_j$ integration and we subtract the final single $n+m$-th order $\epsilon$ pole using standard \MSbar~prescription. The regularized contribution to the double differential rapidity and transverse momentum distribution when the kernel $K$ is inserted $n$-th times on one leg and $m$-th times on the other leg, after the iterative subtraction of the first $n-1$ and $m-1$ collinear singularities turns out to be
\begin{align}
&\frac{d\hat{\sigma}^{n,m}}{dy d\xi_p}\(N,\xi_p,b,\frac{\muf^2}{Q^2},\as\)=\gamma\(N-\frac{i b}{2},\as\)\gamma\(N+\frac{i b}{2},\as\)\notag\\
&\int_0^\infty \frac{d\xi_n}{\xi_n^{1+\epsilon}}\int_0^\infty \frac{d\bar{\xi}_m}{\bar{\xi}_m^{1+\epsilon}} C_{\pt,y}\(N,\xi,\bar{\xi},\xi_p,b,\as\)\times\notag\\
&\times\frac{1}{\(n-1\)!}\frac{1}{\epsilon^{n-1}}\left[\sum_i \frac{\tilde{\gamma}_i\(N-\frac{i b}{2},\as;0\)}{i}\(1-\(\frac{\muf^2}{Q^2\xi_n}\)^{i\epsilon}\frac{\tilde{\gamma}_i\(N-\frac{i b}{2},\as;\epsilon\)}{\tilde{\gamma}_i\(N-\frac{i b}{2},\as;0\)}\)\right]^{n-1}\notag\\
&\times\frac{1}{\(m-1\)!}\frac{1}{\epsilon^{m-1}}\left[\sum_j \frac{\tilde{\gamma}_j\(N+\frac{i b}{2},\as;0\)}{j}\(1-\(\frac{\muf^2}{Q^2\bar{\xi}_m}\)^{j\epsilon}\frac{\tilde{\gamma}_j\(N+\frac{i b}{2},\as;\epsilon\)}{\tilde{\gamma}_j\(N+\frac{i b}{2},\as;0\)}\)\right]^{m-1}
\end{align}
where we introduce the expansion
\beq
\label{eq:gammatilde}
\gamma\(N \pm \frac{i b}{2},\(\frac{\muf^2}{Q^2\xi}\),\as;\epsilon\)=\sum_{i}\tilde{\gamma}_i\(N \pm \frac{i b}{2},\as;\epsilon\)\(\frac{\muf^2}{Q^2\xi}\)^{i \epsilon}.
\eeq

Finally, by summing over all possible insertions and by taking $\epsilon \to 0$, we come to
\begin{align}
\label{eq:doublediffresumfinal}
&\frac{d\hat{\sigma}^{\rm res}}{dy d\xi_p}\(N,\xi_p,b,\frac{\muf^2}{Q^2},\as\)=\gamma\(N-\frac{i b}{2},\as\)\gamma\(N+\frac{i b}{2},\as\) \notag\\
&\qquad\qquad \mathcal{R}\(N-\frac{i b}{2},\as\) \mathcal{R}\(N+\frac{i b}{2},\as\)\notag\\
&\qquad\qquad\int_0^\infty d\xi\, \xi^{\gamma\(N-\frac{i b}{2},\as\)-1}\,\int_0^{\infty} d\bar{\xi}\, \bar{\xi}^{\gamma\(N+\frac{i b}{2},\as\)-1}\,C_{\pt,y}\(N,\xi,\bar{\xi},\xi_p,b,\as\)\notag\\
&\qquad\qquad\times \exp\left[\gamma\(N+\frac{i b}{2},\as\)\ln\frac{Q^2}{\muf^2}\right] \exp\left[\gamma\(N-\frac{i b}{2},\as\)\ln\frac{Q^2}{\muf^2}\right]
\end{align}
where we insert a resummation scheme choice factor $\mathcal{R}$ defined as
\beq
\mathcal{R}\(N,\as\)=\exp\left[-\sum_i \frac{\dot{\tilde{\gamma}}_i\(N,\as\)}{i}\right]
\eeq
with $\dot{\tilde{\gamma}}$ related to the $\epsilon$ expansion of $\tilde{\gamma}\(N,\as;\epsilon\)$, Eq.~\eqref{eq:gammatilde} as
\beq
\tilde{\gamma}\(N,\as;\epsilon\)=\tilde{\gamma}_i\(N,\as\)+\epsilon \dot{\tilde{\gamma}}_i\(N,\as\)+\Ord\(\epsilon^2\).
\eeq
It is important to observe that $\gamma\(N,\as\)=\sum_i \tilde{\gamma}_i\(N,\as\)$ in Eq.~\eqref{eq:doublediffresumfinal} coincides~\cite{Caola:2010kv, Forte:2015gve} with LL$x$ DGLAP resummed anomalous dimension derived by exploiting duality with BFKL kernel. A numerical implementation of this object can be obtained from the public code HELL~\cite{Bonvini:2016wki,Bonvini:2017ogt}. The factor $\mathcal{R}$ is not the only source of dependence from the factorization scheme. A further factor $\mathcal{N}$~\cite{Ciafaloni:2005cg} has to be inserted in order to take into account the non-commutativity between exponentiation and iterative subtraction. By combining these two effects we end up with a unique prefactor $R$ which takes into account all the scheme dependent components. The explicit expression for this factor can be found for example in the original Ref.~\cite{Catani:1994sq}.

Putting $\muf^2=Q^2$ for simplicity we arrive to our final resummed expression for the double differential distribution:
\begin{align}
\label{eq:doublediffresumfinal2}
&\frac{d\hat{\sigma}^{\rm res}}{dy d\xi_p}\(N,\xi_p,b,\frac{\muf^2}{Q^2},\as\)=\gamma\(N-\frac{i b}{2},\as\)\gamma\(N+\frac{i b}{2},\as\) \notag\\
&\qquad\qquad R\(\gamma\(N-\frac{i b}{2},\as\),\as\) R\(\gamma\(N+\frac{i b}{2},\as\),\as\)\notag\\
&\qquad\qquad\int_0^\infty d\xi\, \xi^{\gamma\(N-\frac{i b}{2},\as\)-1}\,\int_0^{\infty} d\bar{\xi}\, \bar{\xi}^{\gamma\(N+\frac{i b}{2},\as\)-1}\,C_{\pt,y}\(N,\xi,\bar{\xi},\xi_p,b,\as\).
\end{align}
Usually in literature, high energy resummed cross sections are expressed using a proper \emph{impact factor}. In this case we can define it as
\begin{align}
\label{eq:impactfactor}
h_{\pt,y}\(N,M_1, M_2, \xi_p,b,\as\)&=M_1 M_2 R\(M_1\) R\(M_2\)\notag\\
&\int_0^\infty d\xi\, \xi^{M_1-1}\,\int_0^{\infty} d\bar{\xi} \bar{\xi}^{M_2-1} C_{\pt,y}\(N,\xi,\bar{\xi},\xi_p,b,\as\).
\end{align}
Therefore, the resummed double differential distribution Eq.~\eqref{eq:doublediffresumfinal2} is obtained from the impact factor as
\beq
\label{eq:relres}
\frac{d\hat{\sigma}^{\rm res}}{dy d\xi_p}\(N,\xi_p,b,\frac{\muf^2}{Q^2},\as\)= h_{\pt,y}\(N,\gamma\(N-\frac{i b}{2},\as\),\gamma\(N+\frac{i b}{2},\as\),\xi_p,b,\as\).
\eeq
Eq.~\eqref{eq:doublediffresumfinal2} and impact factor Eq.~\eqref{eq:impactfactor} represent the main two new results of this paper from the theoretical point of view. They permit the high energy resummation at LL$x$ for double differential distributions. Strictly speaking at LL$x$ we can even discard explicit $N$ dependence in the impact factor, since only the value of it for $N=0$ contributes at this accuracy. However, in the application of the next section we are going to retain such dependence for completeness and we will discard it only when singular behaviour at fixed orders is evaluated.

In conclusion, resummation for rapidity and transverse momentum double differential distribution is obtained by applying Eq.~\eqref{eq:doublediffresumfinal} with $C_{\pt,y}$ in Mellin Fourier space defined as in Eq.~\eqref{eq:Cptyb}. 
%
 We will see all this machinery at work by evaluating LL$x$ behaviour for the Higgs double differential distribution. For simplicity, we are going to limit ourself to the effective field theory case (EFT now on) where Higgs production via gluon fusion is taken as pointlike.

\section{Double differential Higgs boson spectrum at high energy}
\label{sec:Higgshigh}  
This last section will be devoted to the presentation of an explicit application of the theoretical discussion of the previous section, together with some analytic checks. The aim is to verify in a particular process the whole theoretical derivation showed up to now.

We are going to compute the resummation at high energy at LL$x$ of the logs of $\hat{\tau}$, Eq.~\eqref{eq:taudef}, in the double differential cross section with respect to the Higgs transverse momentum and rapidity. As presented in Sec.~\ref{sec:doublediff}, this resummation is performed in Mellin-Fourier space as in Eq.~\eqref{eq:doublediffresumfinal2}.

In the case of Higgs boson production, since final state kinematics in the hard part is a simple $2 \to 1$ case, we prove that residual rapidity dependence in the hard part is subleading. Therefore, the following relation holds
\beq
\label{eq:Cptsinglet}
C_{\pt, y}\(N,\xi,\bar{\xi},\xi_p,b,\as\)=C_{\pt}\(N,\xi,\bar{\xi},\xi_p,\as\)
\eeq
with $C_{\pt}$ the hard transverse momentum distribution for the LO off-shell process
\beq
g^* + g^* \to H.
\eeq
In the case of pointlike Higgs production, $C_{\pt}$ was already evaluated in Ref.~\cite{Forte:2015gve}. Relation Eq.~\eqref{eq:Cptsinglet} brings to another consequence. Resummed expression for the double differential distribution can be obtained directly from the $\pt$-impact factor, defined in Ref.~\cite{Forte:2015gve}, using the following relation
\beq
\label{eq:doublediffresumfinalcoloursinglet}
\frac{d\hat{\sigma}^{\rm res}}{dy d\xi_p}\(N,\xi_p,b,\frac{\muf^2}{Q^2},\as\)= h_{\pt}\(N,\gamma\(N-\frac{i b}{2},\as\),\gamma\(N+\frac{i b}{2},\as\),\xi_p,\as\).
\eeq
The $\pt$-impact factor for the pointlike Higgs boson production case was already been calculated in Ref.~\cite{Forte:2015gve} and turns out to be
\begin{multline}
\label{eq:hpthiggsfinal}
h_{\pt}\(N,M_1,M_2,\xi_p,\as\)=R\(M_1\)R\(M_2\)\sigma_{0}\frac{\xi_p^{M_1+M_2-1}}{\(1+\xi_p\)^N}\\
\left[\frac{\Gamma\(1+M_1\)\Gamma\(1+M_2\)\Gamma\(2-M_1-M_2\)}{\Gamma\(2-M_1\)\Gamma\(2-M_2\)\Gamma\(M_1+M_2\)}\(1+\frac{2M_1M_2}{1-M_1-M_2}\)\right].
\end{multline}

High energy resummation for the rapidity and transverse momentum double differential distribution is now reached; using Eq.~\eqref{eq:doublediffresumfinalcoloursinglet} we immediately obtain the resummed expression for the double differential cross section.
\beq
\label{eq:ggHiggsfinal}
\frac{d\hat{\sigma}_{gg}}{dy d\xi_p}\(N,b,\xi_p,\as\)=h_{\pt}\(N,\gamma\(N-\frac{i b}{2},\as\),\gamma\(N+\frac{i b}{2},\as\),\xi_p,\as\).
\eeq
We insert in previous equation a subscript $gg$ to indicate that resummed expression just obtained is strictly speaking valid only for the gluon-gluon contribution, since we limit ourself to study only the pure gluonic channel at the beginning of our theoretical discussion. Moreover at LL$x$ we can set $N=0$ in the explicit $N$ dependence of the impact factor.

However, all the quark channel components can be derived using the gluon-gluon impact factor~\cite{Harlander:2009my, Forte:2015gve}. Indeed, exploiting known properties of the high energy dynamics~\cite{Harlander:2009my} we obtain the following resummed predictions
\begin{subequations}
\label{eq:Higgsfinalotherchannel}
\begin{align}
\frac{d\hat{\sigma}_{gq}}{dy d\xi_p}\(N,b,\xi_p,\as\)&=\frac{\Cf}{\Ca}\(h_{\pt}\Big(N,\gamma\(N-\frac{i b}{2},\as\),\gamma\(N+\frac{i b}{2},\as\),\xi_p,\as\)\notag\\
&-h_{\pt}\(N,\gamma\(N-\frac{i b}{2},0,\as\),0,\xi_p,\as\)\Big)\\
\frac{d\hat{\sigma}_{qg}}{dy d\xi_p}\(N,b,\xi_p,\as\)&=\frac{\Cf}{\Ca}\Big(h_{\pt}\(N,\gamma\(N-\frac{i b}{2},\as\),\gamma\(N+\frac{i b}{2},\as\),\xi_p,\as\)\notag	\\
& -h_{\pt}\(N,0,\gamma\(N+\frac{i b}{2},\as\),0,\xi_p,\as\)\Big)\\
\frac{d\hat{\sigma}_{qq}}{dy d\xi_p}\(N,b,\xi_p,\as\)&=\(\frac{\Cf}{\Ca}\)^2\Big(h_{\pt}\(N,\gamma\(N-\frac{i b}{2},\as\),\gamma\(N+\frac{i b}{2},\as\),\xi_p,\as\)\notag\\
&-h_{\pt}\(N,0,\gamma\(N+\frac{i b}{2},\as\),0,\xi_p,\as\)\notag\\
&-h_{\pt}\(N,\gamma\(N-\frac{i b}{2},\as\),0,0,\xi_p,\as\)+h_{\pt}\(N,0,0,0,\xi_p,\as\)\Big)
\end{align}
\end{subequations}
for the other channels. With the subscript $q$ we indicate the quark singlet component which is the only component coupled to gluons and then the only one which shows a logarithmic high energy behaviour.

Finally, resummation is now simply achieved by evaluating LL$x$ anomalous dimension $\gamma\(N,\as\)$, and by performing inverse Fourier-Mellin transform. A particularly stable implementation for $\gamma$ is the one presented in Refs.~\cite{Bonvini:2016wki,Bonvini:2017ogt} and implemented in the code HELL.

Let us now conclude this section with some consistency check. The integration over $\xi_p$ of Eq.~\eqref{eq:ggHiggsfinal} or Eqs.~\eqref{eq:Higgsfinalotherchannel} returns the same predictions of Ref.~\cite{Caola:2010kv} for the resummation of the Higgs rapidity distribution; integration over $y$ (achieved in Fourier space setting $b=0$) reconstructs the known resummation of transverse momentum distribution of Ref.~\cite{Forte:2015gve} and, finally, integration over both $\xi_p$ and $y$ is consistent with known resummation of Ref.~\cite{Hautmann:2002tu} for the Higgs inclusive cross section.

To further prove the correctness of our derivation, we will not focus only on the resummed result but we are going to expand Eq.~\eqref{eq:ggHiggsfinal} in power of $\as$ and to crosscheck first coefficients against fixed order evaluations. The technique we will use to perform this check is similar to the one presented in Ref.~\cite{Muselli:2017ikh}. 

We expand our result in powers of $\as$ using following equalities
\begin{align}
R\(M\)&=1+\frac{8}{3}\zeta_3 M^3+ \Ord\(M^4\)\\
\gamma\(N,\as\)&=\frac{\Ca}{\pi} \frac{\as}{N}+\Ord\(\as^4\),
\end{align} 
obtaining:
\beq
\frac{d\hat{\sigma}_{gg}}{dy d\xi_p}\(N,b,\xi_p,\as\)=\sigma_0\sum_{k=0}^{\infty}\as^k C_k\(N,b, \xi_p\)
\eeq
with
\begin{subequations}
\label{eq:coefficients}
\begin{align}
C_0\(N,b,\xi_p\)&=\delta\(\xi_p\),\\
C_1\(N,b,\xi_p\)&=\frac{\Ca}{\pi}\plusq{\frac{1}{\xi_p}}\(\frac{1}{N+\frac{i b}{2}}+\frac{1}{N-\frac{i b}{2}}\),\\
\label{eq:C2Higgs}
C_2\(N,b,\xi_p\)&=\(\frac{\Ca}{\pi}\)^2\( \plusq{\frac{\ln\xi_p}{\xi_p}}\(\frac{1}{N+\frac{i b}{2}}+\frac{1}{N-\frac{i b}{2}}\)^2+\frac{1}{N-\frac{i b}{2}} \frac{1}{N+\frac{i b}{2}} \delta\(\xi_p\)\).
\end{align}
\end{subequations}
In performing this expansion,  we set $N=0$ in the explicit impact factor $N$ dependence to obtain a pure LL$x$ result. Moreover, in Eqs.~\eqref{eq:coefficients}, we define plus distributions as
\beq
\label{eq:plusxidef}
\int_0^{\xi_{\rm max}} d\xi_p\, \plusq{f\(\xi_p\)} g\(\xi_p\)=\int_0^{\xi_{\rm max}}d\xi_p\, f\(\xi_p\)\(g\(\xi_p\)-\Theta\(1-\xi_p\) g\(0\)\)
\eeq
with $\xi_{\rm max}=\frac{\(1-\hat{\tau}\)^2}{4\hat{\tau}}\approx \infty$ the $\xi_p$ upper kinematic limit and $\Theta$ the Heaviside distribution.

We are going to limit ourselves to the gluon-gluon case for simplicity; the interested reader can check also the other channels with similar techniques.

Comparisons with the exact fixed order evaluations for the first coefficients $C_1$ and $C_2$ will be performed in the next subsection. 

\subsection{Check against fixed order evaluation}
\label{subsec:check}
In this subsection we want to check the leading log limit of the double differential distribution with respect to rapidity and transverse momentum for the Higgs boson production process in gluon fusion. We will follow analogue comparison performed in Ref.~\cite{Muselli:2017ikh} and we are going to compute the high energy limit of the first perturbative orders by explicit computation of relevant Feynman Diagrams. For simplicity, we will use as independent variables $\hat{\tau}$, $\xi_p$ and $u=e^{-2y}$, hence evaluating $\frac{d\hat{\sigma}}{du d\xi_p}$ rather than $\frac{d\hat{\sigma}}{dy d\xi_p}$. We will come back to the set $\hat{\tau}$, $\xi_p$ and $y$ at the end of the calculus, right before the computation of the Mellin-Fourier transform.

We start from the single emission. Diagrams contributing at LL$x$ are drawn in Fig.~\ref{fig:diagram1}. With the red circle we indicate the \MSbar~collinear counterterm, which occurs to make the sum finite. In computing the square modulus of the amplitude we can ignore interferences between diagrams since they are subleading in the high energy limit.

\begin{table}[htb!]
\centering
\caption{NLO Feynman Diagrams and \MSbar~subtraction}
\begin{tabular} {c c}
\includegraphics[width=0.5\linewidth]{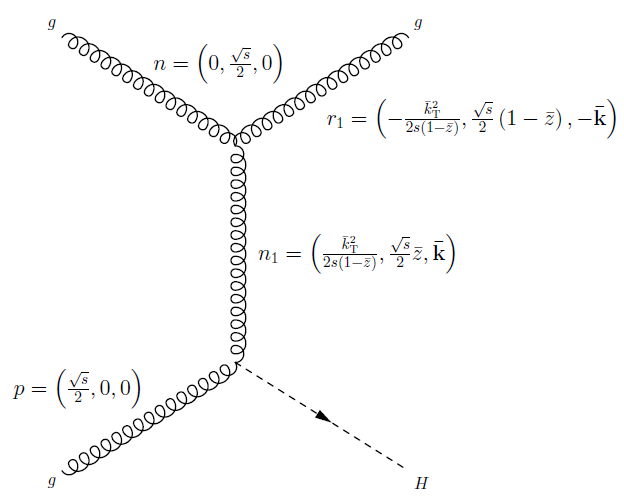} &
\includegraphics[width=0.35\linewidth]{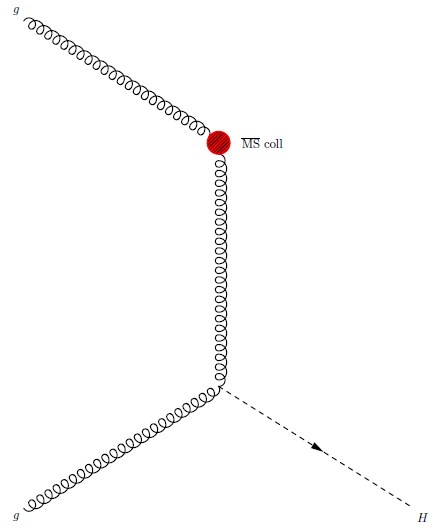} \\
\includegraphics[width=0.5\linewidth]{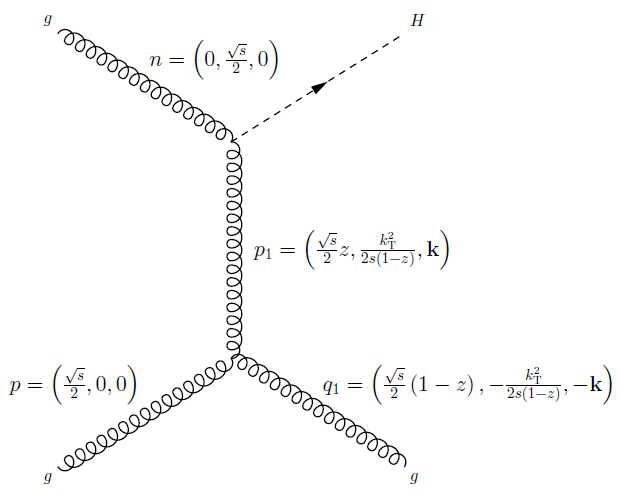} &
\includegraphics[width=0.35\linewidth]{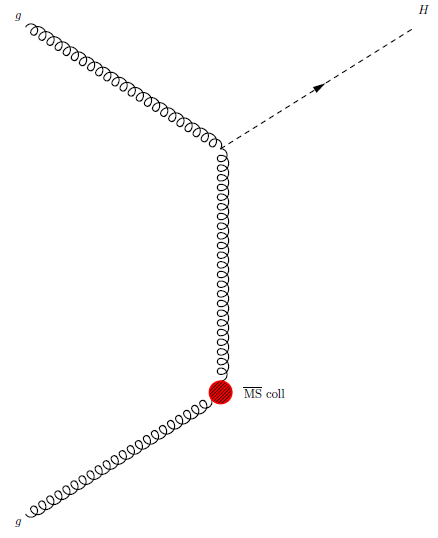}
\end{tabular}
\label{fig:diagram1}
\end{table}

For the diagrams in the first row of Fig.~\ref{fig:diagram1}, the associate double differential cross section can be written as
\begin{align}
\label{eq:singularemission}
\frac{d\sigma_1}{d\xi_p du}=\frac{1}{2u}\frac{d\sigma_1}{d\xi_p dy}&=\sigma_0 \frac{\bar{z}}{\hat{\tau}}\delta\(1-\frac{\bar{z}}{\hat{\tau}}+\bar{\xi}\)\left[\bar{\alpha}_s\frac{d\bar{z}}{\bar{z}}\frac{d\bar{\xi}}{\bar{\xi}^{1+\epsilon}}\right]\delta\(\xi-\xi_p\)\delta\(u-\bar{z}\)\notag\\
&+\frac{\sigma_0 \bar{\alpha}_s}{\epsilon}\delta\(\xi_p\)\delta\(u-\hat{\tau}\).
\end{align}
where integrations over $\bar{z}$, $\bar{\xi}$ are implicitly assumed. We are going to use the bar to indicate all the Sudakov components related to the upper leg of the diagrams of Fig.~\ref{fig:diagram1}. Moreover we introduce
\beq
\bar{\alpha}_s=\frac{\as\(\mu^2\)\mu^{2\epsilon} \Ca}{\pi}
\eeq
with $\epsilon$ the dimensional regulator, and we ignore all the $\frac{\(4 \pi\)^{\epsilon}}{\Gamma\(1-\epsilon\)}$ terms which are systematically subtracted order by order in \MSbar~scheme.

Kinematic limits for $\hat{\tau},\xi_p,u,\bar{z},\bar{\xi}$ in the high energy regime are the following
\begin{align}
&0<\hat{\tau}<1, & &\xi_p>0, & &0<u<1, & &0<\bar{z}<1, & &\bar{\xi}>0.
\end{align}
It is important to note that, due to the presence of the collinear counterterm, Eq.~\eqref{eq:singularemission} is finite in $d=4$ dimension.
Now to come to the final result for the single emission we perform the integration over $\bar{\xi}$ and $\bar{z}$ using the two delta constraints. We thus obtain
\beq
\label{eq:single}
\frac{d\sigma_1}{d\xi_p du}=\sigma_0 \bar{\alpha}_s\left[\frac{1}{\xi_p^{1+\epsilon}} \delta\(u-\hat{\tau}\(1+\xi_p\)\)+\frac{1}{\epsilon}\delta\(\xi_p\)\delta\(u-\hat{\tau}\)\right].
\eeq
with now $\hat{\tau},\xi_p,u$ running on the following ranges
\begin{align}
&0<\hat{\tau}<1, & &0<\xi_p<\frac{1-\hat{\tau}}{\hat{\tau}}, & &0<u<1.
\end{align}
In the high energy limit at LL$x$, two further simplifications on Eq.~\eqref{eq:single} occur. First, the delta constraint on the rapidity can be simplified according to
\beq
\label{eq:deltauhigh}
\delta\(u-\hat{\tau}\(1+\xi_p\)\) \approx \delta\(u-\hat{\tau}\);
\eeq
then, the $\xi_p$ upper limit can be confused with infinity since $\hat{\tau} \to 0$.

The last step we need to carry on is the $\epsilon$ expansion. Using the following expansion into Eq.~\eqref{eq:single}
\begin{align}
\label{eq:expansion}
&\frac{1}{\xi_p^{1+\epsilon}}=-\frac{1}{\epsilon}\delta\(\xi_p\)+\sum_{j=0}^\infty \frac{(-1)^j \epsilon^j}{j!} \plusq{\frac{\ln^j \xi_p}{\xi_p}}
\end{align}
with plus distribution defined as in Eq.~\eqref{eq:plusxidef}, we come to
\beq
\label{eq:finalup}
\frac{d\sigma_1}{d\xi_p du}=\sigma_0\, \bar{\alpha}_s \plusq{\frac{1}{\xi_p}}\delta\(u-\hat{\tau}\).
\eeq
which is our final result. To obtain the complete NLO correction for the EFT Higgs boson production at high energy we need to compute the remaining diagrams of Fig.~\ref{fig:diagram1}. They are obtained from Eq.~\eqref{eq:finalup} by performing the replacement $y \to -y$ which means $u \to \frac{1}{u}$. Therefore the final NLO result turns out to be:
\begin{align}
\label{eq:NLO}
\frac{d\sigma_1}{d\xi_p du}&=\sigma_0\, \bar{\alpha}_s \plusq{\frac{1}{\xi_p}}\(\delta\(u-\hat{\tau}\)+\delta\(\frac{1}{u}-\hat{\tau}\)\)
\end{align}

We now move to the NNLO order. The diagrams contributing at this order are collected in Fig.~\ref{fig:diagram2} together with the proper collinear subtractions.

\begin{flushleft}
\begin{table}[hb!]
\caption{NNLO Feynman Diagrams and \MSbar~subtraction}
\begin{tabular} {c c c c}
\vspace{5mm}
\includegraphics[width=0.2\linewidth]{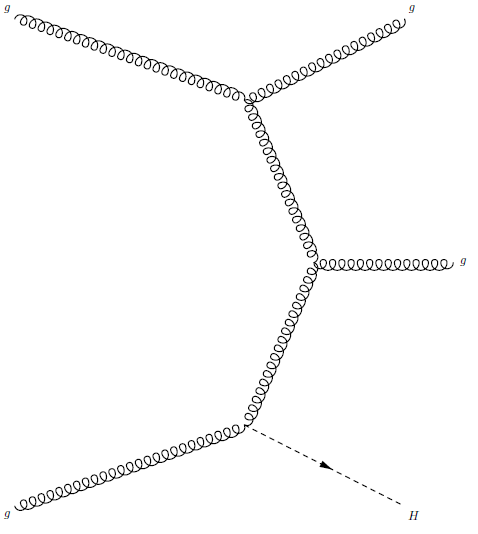} &
\includegraphics[width=0.2\linewidth]{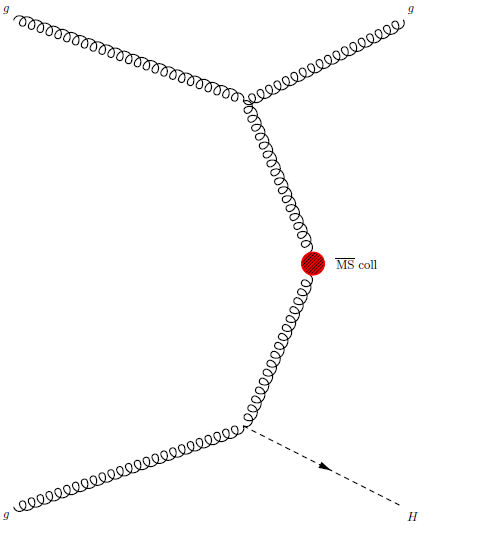} &
\includegraphics[width=0.2\linewidth]{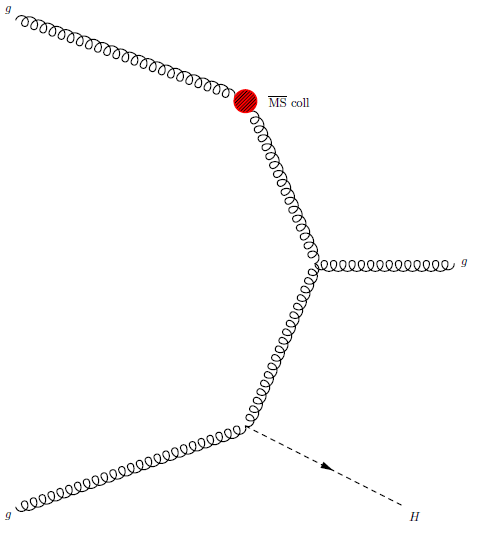} &
\includegraphics[width=0.2\linewidth]{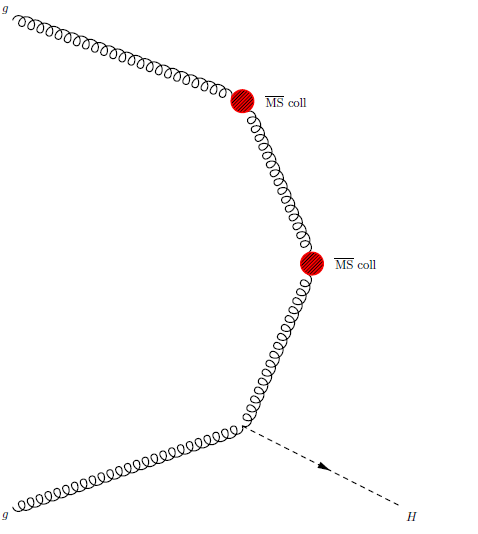} \\
\includegraphics[width=0.2\linewidth]{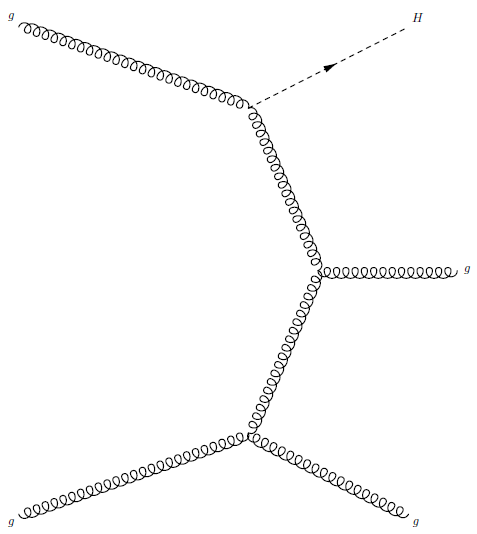} &
\includegraphics[width=0.2\linewidth]{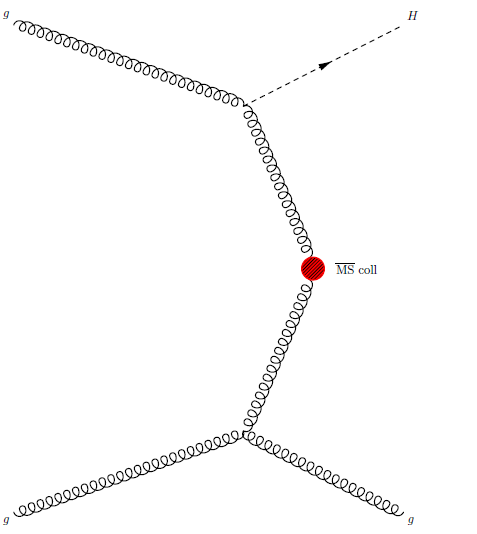} &
\includegraphics[width=0.2\linewidth]{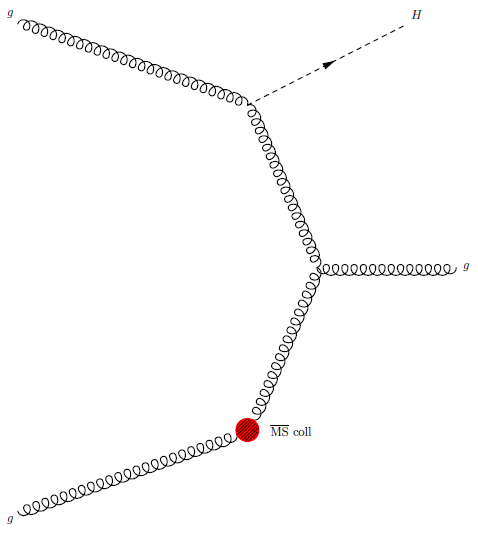} &
\includegraphics[width=0.2\linewidth]{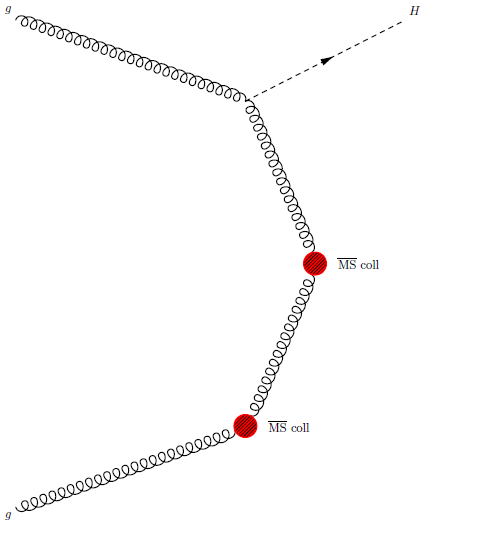} \\
\includegraphics[width=0.2\linewidth]{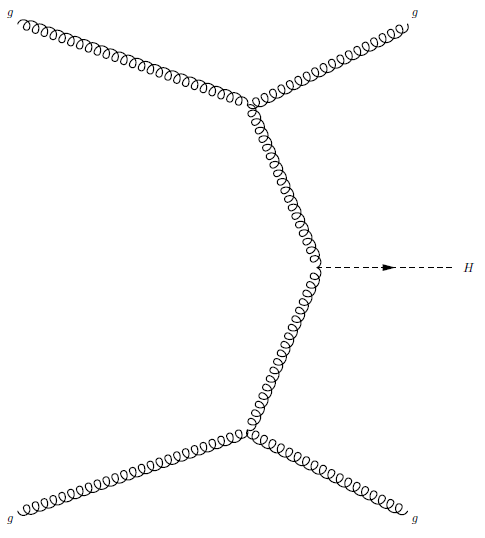} &
\includegraphics[width=0.2\linewidth]{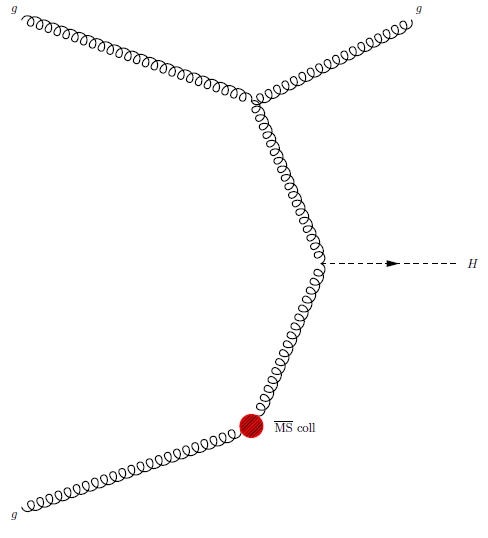} &
\includegraphics[width=0.2\linewidth]{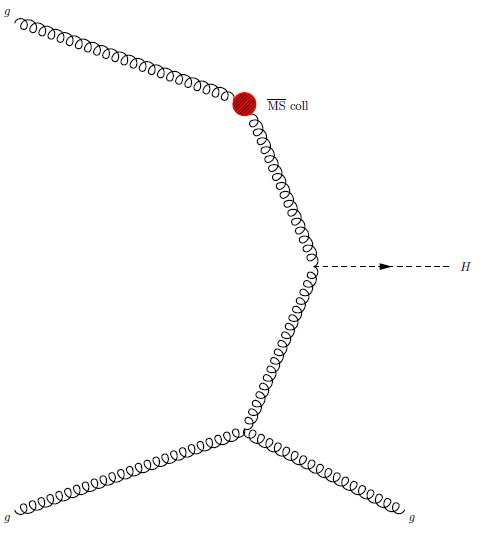} &
\includegraphics[width=0.2\linewidth]{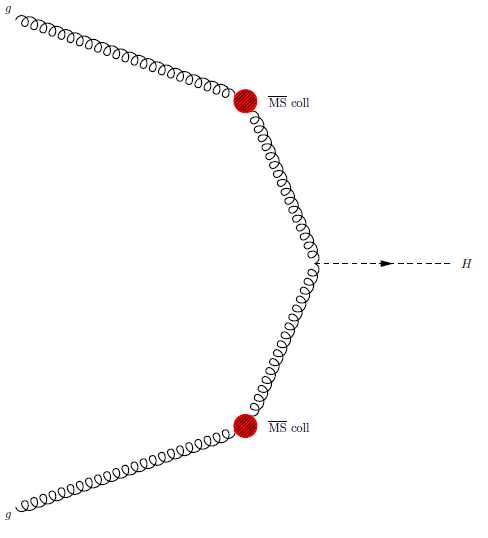}
\end{tabular}
\label{fig:diagram2}
\end{table}
\end{flushleft}

We start our computation from the diagrams of the first row. As before the contribution from the second row of diagrams is obtained from the first one by the substitution $u \to \frac{1}{u}$. The double differential cross section in $d=4-2\epsilon$ dimensions can be written as
\begin{align}
\label{eq:doubleemission}
\frac{d\sigma_2}{d\xi_p du}&=\sigma_0 \frac{\bar{z}_1 \bar{z}_2}{\hat{\tau}}\delta\(1-\frac{\bar{z}_1\bar{z}_2}{\hat{\tau}}+\bar{\xi}_1\)\left[\bar{\alpha}_s\frac{d\bar{z}_2}{\bar{z}_2}\frac{d\bar{\xi}_2}{\bar{\xi}_2^{1+\epsilon}}\right]\left[\bar{\alpha}_s \frac{d\bar{z}_1}{\bar{z}_1}\frac{d\bar{\xi}_1}{\bar{\xi}_1^{1+\epsilon}}\right]\notag\\
&\qquad\qquad\qquad\delta\(\bar{\xi}_1-\xi_p\)\delta\(u-\bar{z}_1\bar{z}_2\)\notag\\
&+\sigma_0\, \bar{\alpha}_s^2\left[\frac{1}{2\epsilon^2}\delta\(\xi_p\)\delta\(u-\hat{\tau}\)+\frac{1}{\epsilon}\left[\frac{1}{\xi_p^{1+\epsilon}}\delta\(u-\hat{\tau}\(1+\xi_p\)\)\right]\right].
\end{align}
with the following limits of integration at high energy for the various quantities
\begin{align}
&0<\hat{\tau}<1, & &\xi_p>0, & &0<u<1, & &0<\bar{z}_1,\bar{z}_2<1, & &\bar{\xi}_1>0, & &0<\bar{\xi}_2 <\bar{\xi}_1.
\end{align}
Now we solve the integration over $\bar{\xi}_1$, $\bar{z}_1$ using the delta constraints contained in Eq.~\eqref{eq:doubleemission}. We thus obtain:
\begin{align}
\frac{d\sigma_2}{d\xi_p du}&=\sigma_0 \bar{\alpha}_s^2\left[\frac{d\bar{z}_2}{\bar{z}_2} \frac{d\bar{\xi}_2}{\bar{\xi}_2^{1+\epsilon}}\right]\frac{1}{\xi_p^{1+\epsilon}}\delta\(u-\hat{\tau}\(1+\xi_p\)\)\notag\\
&+\sigma_0\, \bar{\alpha}_s^2\left[\frac{1}{2\epsilon^2}\delta\(\xi_p\)\delta\(u-\hat{\tau}\)+\frac{1}{\epsilon}\left[\frac{1}{\xi_p^{1+\epsilon}}\delta\(u-\hat{\tau}\(1+\xi_p\)\)\right]\right]
\end{align}
with now the integration limits given by
\begin{align}
&0<\hat{\tau}<1, & &0<\xi_p<\frac{1-\hat{\tau}}{\hat{\tau}}\approx \infty, & &0<u<1, & &u< \bar{z}_2<1, & &0<\bar{\xi}_2 < \xi_p.
\end{align}
By performing the integrations over $\bar{z}_2$ and $\bar{\xi}_2$ we obtain
\begin{align}
\frac{d\sigma_2}{d\xi_p du}&=\sigma_0\, \bar{\alpha}_s^2 \ln\frac{1}{u}\left[-\frac{1}{\epsilon}\frac{1}{\xi_p^{1+2\epsilon}}\delta\(u-\hat{\tau}\(1+\xi_p\)\)\right]\notag\\
&+\sigma_0\, \bar{\alpha}_s^2\left[\frac{1}{2\epsilon^2}\delta\(\xi_p\)\delta\(u-\hat{\tau}\)+\frac{1}{\epsilon}\left[\frac{1}{\xi_p^{1+\epsilon}}\delta\(u-\hat{\tau}\(1+\xi_p\)\)\right]\right].
\end{align}
The last step is to use equality Eq.~\eqref{eq:expansion} twice to expand the result around $\epsilon=0$ and to simplify delta constraint according to Eq.~\eqref{eq:deltauhigh}. We write
\beq
\label{eq:NNLO1}
\frac{d\sigma_2}{d\xi_p du}=\sigma_0\, \bar{\alpha}_s^2 \ln\frac{1}{u}
\plusq{\frac{\ln\xi_p}{\xi_p}}\delta\(u-\hat{\tau}\)
\eeq
which is the final contribution to the double differential cross section from the diagrams of the first row of Fig.~\ref{fig:diagram2}. As said before, the contribution from the second row is obtained from Eq.~\eqref{eq:NNLO1} by performing the substitution $u \to \frac{1}{u}$.

Then, we have to evaluate graphs of the third row of Fig.~\ref{fig:diagram2}. Their contribution to the double differential cross section in $d=4-2\epsilon$ dimensions at high energy turns out to be:
\begin{align}
\label{eq:NNLOtwoleg}
\frac{d\sigma_2}{d\xi_p du}&=\sigma_0 \frac{z \bar{z}}{\hat{\tau}}\delta\(1-\frac{z \bar{z}}{\hat{\tau}}+\xi_p\)\left[\bar{\alpha}_s \frac{dz}{z} \frac{d \xi}{\xi^{1+\epsilon}}\right]\left[\bar{\alpha}_s \frac{d\bar{z}}{\bar{z}} \frac{d\bar{\xi}}{\bar{\xi}^{1+\epsilon}}\right]\cos^2\theta\frac{d\theta}{2\pi}\notag\\
&\qquad\qquad\delta\(\xi_p-\xi-\bar{\xi}-2\sqrt{\xi \bar{\xi}}\cos\theta\)\delta\(u-\frac{\bar{z}}{z}\)\notag\\
&+\frac{\sigma_0\,\bar{\alpha}_s^2}{u}\left[\frac{1}{\epsilon}\frac{1}{\xi_p^{1+\epsilon}}+\frac{1}{2\epsilon^2}\delta\(\xi_p\)\right]
\end{align}
The integration domain is composed by two disjointed regions
\begin{align}
&0<\hat{\tau}<1, & & & &0<\hat{\tau}<1\notag\\
&\hat{\tau}<u<1, & &\bigcup & &1<u<\frac{1}{\hat{\tau}}\\
&0<\xi_p<\frac{u}{\hat{\tau}}-1, & & & &0<\xi_p<\frac{1}{u \hat{\tau}}-1\notag.
\end{align}
and $\bar{\xi},\xi>0$, $0<z,\bar{z}<1$, $0<\theta<2\pi$.
We are going to compute next steps limiting ourself to the first region. The complete result is obtained by symmetrizing the final expression with respect to the transformation $u \to \frac{1}{u}$. Moreover, Eq.~\eqref{eq:NNLOtwoleg} is invariant under exchange of $\xi$ and $\bar{\xi}$. We can thus halve the integration region by requiring $\xi > \bar{\xi}$ and recover the other part by exploiting this symmetry. We then perform the following change of variables:
\begin{align}
\xi&=\xi_p\,\xi_1, & \bar{\xi}&=\xi_p\,w\,\xi_1, &\cos\theta=t,
\end{align}
with
\begin{align}
&\xi_1>0, & &0<w<1, & &-1<t<1.
\end{align}
Using this new set of integration variables and using delta constraints to eliminate integrations over $z$ and $\bar{z}$, we rewrite Eq.~\eqref{eq:NNLOtwoleg} as:
\begin{align}
\frac{d\sigma_2}{d\xi_p du}&=\frac{2\sigma_0\, \bar{\alpha}_s^2}{\pi u}\frac{1}{\xi_p^{1+2\epsilon}}\frac{dw}{w^{1+\epsilon}}\frac{d\xi_1}{\xi_1^{1+2\epsilon}} \frac{t^2 dt}{\sqrt{1-t^2}}\delta\(1-\xi_1\(1+w+2\sqrt{w}t\)\)\notag\\
&+\frac{\sigma_0\,\bar{\alpha}_s^2}{u}\left[\frac{1}{\epsilon}\frac{1}{\xi_p^{1+\epsilon}}+\frac{1}{2\epsilon^2}\delta\(\xi_p\)\right].
\end{align}
Next step is to use last delta constraint to solve integration over $\xi_1$. Performing the integrations over $w$ and $t$ we obtain:
\begin{align}
\frac{d\sigma_2}{d\xi_p du}&=\frac{\sigma_0\, \bar{\alpha}_s^2}{\pi u}\frac{1}{\xi_p^{1+2\epsilon}}\left[\int_{-1}^1 \frac{2 t^2 dt}{\sqrt{1-t^2}}\int_0^1 dw\frac{\(1+w+2\sqrt{w}t\)^{2\epsilon}-1}{w^{1+\epsilon}}-\frac{\pi}{\epsilon}\right]\notag\\
&+\frac{\sigma_0\,\bar{\alpha}_s^2}{u}\left[\frac{1}{\epsilon}\frac{1}{\xi_p^{1+\epsilon}}+\frac{1}{2\epsilon^2}\delta\(\xi_p\)\right]\\
&=\frac{\sigma_0\, \bar{\alpha}_s^2}{u}\frac{1}{\xi_p^{1+2\epsilon}}\left[-\frac{1}{\epsilon}-\epsilon+\Ord\(\epsilon^2\)\right]+\frac{\sigma_0\,\bar{\alpha}_s^2}{u}\left[\frac{1}{\epsilon}\frac{1}{\xi_p^{1+\epsilon}}+\frac{1}{2\epsilon^2}\delta\(\xi_p\)\right]
\end{align}
Finally we come to the desired result by expanding in $\epsilon=0$ using relation Eq.~\eqref{eq:expansion}\footnote{Note that in high energy limit even in this case $\xi_p$ upper limit can be confused with $\infty$}. We obtain (recovering also the other region):
\begin{align}
\label{eq:NNLO2}
\frac{d\sigma_2}{d\xi_p du}&=\sigma_0\, \bar{\alpha}_s^2\left\{\left[\left(\plusq{\frac{\ln\xi_p}{\xi_p}}+\frac{1}{2}\delta\(\xi_p\)\right)\frac{\Theta\(1-u\)}{u}\right]+\left[u \leftrightarrow \frac{1}{u}\right]\right\}.
\end{align}
Performing the sum we obtain the full NNLO:
\begin{align}
\label{eq:NNLO}
\frac{d\sigma_2}{d\xi_p du}&=\sigma_0\, \bar{\alpha}_s^2\Bigg\{\Bigg[\ln\frac{1}{u}\plusq{\frac{\ln\xi_p}{\xi_p}}\delta\(u-\hat{\tau}\)+\frac{1}{u}\plusq{\frac{\ln\xi_p}{\xi_p}}\notag\\
&+\frac{1}{2u}\delta\(\xi_p\)\Bigg]\Theta\(1-u\)+\left[u \leftrightarrow \frac{1}{u}\right]\Bigg\}.
\end{align}

Eq.~\eqref{eq:NLO} and Eq.~\eqref{eq:NNLO} represent the high energy limit of the NLO and NNLO respectively in momentum space. By taking Mellin-Fourier transform with respect to $\hat{\tau}$ and $y$, they need to coincide with $C_1$ and $C_2$, Eqs.~\eqref{eq:coefficients}.

Fourier transform with respect to $y$ can be rewritten in terms of $u$ as
\beq
\int_{-\ln x}^{\ln x} dy\,e^{-i b y} \frac{d\sigma}{dy d\xi_p}=\int_{\frac{1}{x}}^x du\,u^{-\frac{i b}{2}} \frac{1}{2u} \frac{d\sigma}{dy d\xi_p}=\int_{\frac{1}{x}}^x du\,u^{-\frac{i b}{2}} \frac{d\sigma}{du d\xi_p}.
\eeq
With straightforward calculations, we obtain in Mellin-Fourier space
\beq
\frac{d\sigma_1}{d\xi_p dy}\(N,b,\xi_p,\as\)=\sigma_0\frac{\Ca\as}{\pi}\plusq{\frac{1}{\xi_p}}\(\frac{1}{N-\frac{i b}{2}}+\frac{1}{N+\frac{i b}{2}}\)
\eeq
for the NLO and
\begin{multline}
\frac{d\sigma_2}{d\xi_p dy}\(N,b,\xi_p,\as\)=\sigma_0 \(\frac{\Ca\as}{\pi}\)^2\\
\left\{\plusq{\frac{\ln\xi_p}{\xi_p}}\(\frac{1}{N-\frac{i b}{2}}+\frac{1}{N+\frac{i b}{2}}\)^2+\frac{1}{N^2+\frac{b^2}{4}}\delta\(\xi_p\)\right\}
\end{multline}
for the NNLO. Final results in Mellin-Fourier space are in perfect agreement with our predictions Eqs.~\eqref{eq:coefficients}, giving in this way a strong cross-check on the whole construction.

Moreover, our final results permit also to check expansions at first orders for the resummed single rapidity or transverse momentum distribution of Refs.~\cite{Caola:2010kv, Forte:2015gve}, or for the resummed total cross section of Ref.~\cite{Hautmann:2002tu}. Indeed by integrating over $y$ - thus setting $b=0$ in Fourier space - we recover high energy behaviour of transverse momentum distribution~\cite{Forte:2015gve}; while performing integration over $\xi_p$ we obtain
\begin{align}
\frac{d\sigma_1}{d\xi_p dy}\(N,b,\as\)&=\sigma_0\, \as \frac{\Ca}{\pi}\frac{1}{N}\(\frac{1}{N-\frac{i b}{2}}+\frac{1}{N+\frac{i b}{2}}\) +\Ord\(\frac{1}{N}\) \\
\frac{d\sigma_2}{d\xi_p dy}\(N,b,\xi_p,\as\)&=\sigma_0\, \as^2 \(\frac{\Ca}{\pi}\)^2\frac{1}{N^2}\(\frac{1}{N-\frac{i b}{2}}+\frac{1}{N+\frac{i b}{2}}\)^2+\Ord\(\frac{1}{N^3}\)
\end{align} 
which agrees with predictions of Ref.~\cite{Caola:2010kv}. It is interesting to note that the second term of Eq.~\eqref{eq:NNLO} (or Eq.~\eqref{eq:C2Higgs}) which is not subleading at fixed $\xi_p$, becomes subleading after $\xi_p$ integration. This is due to the pointlike nature of the effective interaction.

Finally, complete integration over $y$ and $\xi_p$ clearly recover known results for total cross section.

\section{Conclusions and Outlooks}
\label{sec:concl}
Summarizing, in this paper we have presented general high energy resummation for any double differential rapidity and transverse momentum distributions. This resummation at LL$x$ is performed in Mellin-Fourier space in order to factorize rapidity dependence. Final results, Eq.~\eqref{eq:ggHiggsfinal} and~\eqref{eq:Higgsfinalotherchannel} (or the associated \emph{impact factors} Eq.~\eqref{eq:impactfactor}) are closely related to the analogue ones for single rapidity distribution presented in Ref.~\cite{Caola:2010kv} and for single transverse momentum distribution presented in Ref.~\cite{Forte:2015gve}. The theory just exposed is completely general and it can be applied indifferently to colourless or coloured final state. In the second case, however, corrections, due to gauge-invariance or indistinguishability of hard part final state, may be necessary~\cite{Zoia:2017vfn}.

As a cross-check, we evaluate resummed expressions in the case of pointlike Higgs boson production. Due to its simple kinematics, in this case, we are able to use the calculation already shown in Ref.~\cite{Forte:2015gve} to compute the hard part.  We prove that final resummed expressions are consistent upon suitable integrations with known high energy resummed predictions of Refs.~\cite{Caola:2010kv,Forte:2015gve, Hautmann:2002tu} for rapidity distribution, transverse momentum distribution, and inclusive cross section, respectively. Moreover, we expand up to NNLO our resummed result and we cross-check the prediction against fixed-order evaluations.

This paper represents a sort of conclusion of the research project started years ago with the first extension of Ref.~\cite{Caola:2010kv} of high energy resummation at LL$x$ to more exclusive observables. However, it is only the first step toward a almost fully exclusive description of the high energy dynamics; this knowledge is necessary, as pointed out in Ref.~\cite{Zoia:2017vfn}, for many applications, especially in the context of jet physics.

Another future outlook of this research is its numerical implementation, at least for simple processes as the Higgs boson or Drell-Yan pair production. Very recently, a general public code HELL was implemented to deal with high energy resummation first of DGLAP evolution and then of inclusive cross section~\cite{Bonvini:2016wki,Bonvini:2017ogt}. Differential high energy resummation represents the next step and phenomenological resummed predictions are in preparation. 

\acknowledgments{We are grateful to Simone Marzani and Stefano Forte for useful discussions about Fourier transform and single rapidity high energy resummation.}

\end{document}